# Capacitance-voltage measurements of $(Bi_{1-x}Sb_x)_2Te_3$ field effect devices

*Jimin Wang*[1,*], *Markus Schitko*[1], *Gregor Mussler*[2,3], *Detlev Grützmacher*[2,3], *and Dieter Weiss*[1,*]

[1]*Institute for Experimental and Applied Physics, University of Regensburg, D-93040 Regensburg, Germany*

[2]*Peter Grünberg Institute (PGI-9) and JARA-Fundamentals of Future Information Technology, Forschungszentrum Jülich GmbH, 52425 Jülich, Germany*

[3]*Helmholtz Virtual Institute for Topological Insulators (VITI), Forschungszentrum Jülich, 52425 Jülich, Germany*

Capacitance-voltage (*C-V*) traces in n-type-$(Bi_{1-x}Sb_x)_2Te_3$/oxide/metal capacitor structures using an AC capacitance bridge are investigated. By tuning the top gate voltage from positive to negative values, the system at the interface is tuned from accumulation, via depletion into inversion. Our results show the typical low-frequency and high frequency *C-V* traces, depending on measuring frequency, temperature and illumination intensity and reflecting their sensitive dependence on recombination/generation rates. Superimposed a strong hysteresis under inversion is also observed which is ascribed to the presence of conventional localized surface states which coexist with topological surface states.

## 1. Introduction

Topological insulators (TIs) are a new class of materials with unique electronic properties featuring conductive surface states and insulating bulk. [1, 2] HgTe and binary compounds like

Bi$_2$Se$_3$, Bi$_2$Te$_3$, and Sb$_2$Te$_3$, e.g., are typical TIs. [3] The characteristic surface states in TIs form at the interface between materials with different topological index. In many cases, however, the bulk is not insulating due to intrinsic defects, causing strong *n*- or *p*-doping, so that bulk conduction prevails. For example, pure Bi$_2$Te$_3$, is *n*-type with typical carrier concentrations *n* on the order of 10$^{19}$ cm$^{-3}$, pure Sb$_2$Te$_3$ is *p*-type having a hole concentration *p* in the same range. One way to suppress bulk conduction is compensation, which is achieved by combining *n*- and *p*- type materials like Bi$_2$Te$_3$ and Sb$_2$Te$_3$ to form (Bi$_{1-x}$Sb$_x$)$_2$Te$_3$ (BST). In BST the carrier type can be tuned from *n* to *p*-type by varying the Sb concentration (*x* value). At the transition point, the carrier concentration is typically in the range of 10$^{18}$ cm$^{-3}$ or even lower, associated with a correspondingly lower bulk conduction. [4, 5] The observation of a well-resolved quantum Hall effect in BST and similar compensated TIs shows the feasibility of this method. [6] Here, we focus on capacitance-voltage (*C-V*) measurements, being a standard tool to investigate metal-oxide-semiconductor (MOS) devices. [7] For example, *C-V* traces give information about interface states or charges in the oxides. In TIs we can envision three scenarios: (i) In case of high, metallic like bulk conduction we expect the capacitance (per unit area) of metal-insulator-TI structures to be essentially constant and given by $C_0 = \varepsilon\varepsilon_0/d$ with $\varepsilon_0$ the dielectric constant of vacuum and $\varepsilon$, *d* the relative dielectric constant, and thickness of the insulator, respectively. (ii) In case of pure surface conduction (truly insulating bulk) the measured capacitance $C$ reflects the quantum capacitance $e^2D$, with *e* the elementary charge and *D* the thermodynamic density of states at the Fermi level $E_F$, $C^{-1} = C_0^{-1} + (e^2D)^{-1}$. This scenario has also been explored, for example, in the bulk insulating TI HgTe. [8] (iii) In case of intermediate bulk conduction, one might expect that the interface can be tuned by the field effect from accumulation via depletion to inversion, and a mixture of quantum capacitance and MOS field effect capacitance depending on their respective contributions will be obtained. [7, 9] The situation explored below is in the limit of a dominating MOS field effect capacitance.

## 2. Experimental section

BST films with thickness $d$ between 8 to 12.5 nm were grown on silicon on insulator substrates by molecular beam epitaxy. Samples with a nominal Sb concentration of around 42% were chosen as they displayed the lowest carrier concentration ($10^{18}$ ~$10^{19}$ cm$^{-3}$). [5] A typical device used both for transport and capacitance measurements is shown in the inset of Figure 1(a). A Hall-bar mesa was defined by optical lithography, followed by Ar ion milling. The area of the insulator was then defined by electron beam lithography. In the following, measurements of three BST devices are discussed. BST1 and BST2 are nominally identical, having thickness of 12.5 nm. The capacitor consists of 20 nm SiO$_2$ plus 50 nm Si$_3$N$_4$ grown by plasma enhanced chemical vapor deposition. After that, 10 nm Ti and 100 nm Au films were deposited by electron beam and thermal evaporation to form gate and Ohmic contacts. Device BST3 ($d \sim 8$ nm), used for light controlled experiments, was capped in the MBE chamber with 0.7 nm of Al, which was subsequently oxidized, and further capped with 70 nm Al$_2$O$_3$, grown by atomic layer deposition (ALD). A 7 nm thick semi-transparent NiCr gate contact was deposited for easier transmission of the light. [10] The transport measurements for all devices were performed in an Oxford cryostat at temperatures down to 1.4 K and applying magnetic fields up to 1 T. The capacitance-voltage measurements were done with an AH2700 capacitance bridge by mixing a DC top gate voltage ($V_{tg}$) and a modulation AC voltage ($V_{AC}$) of 0.5 V at frequencies between 50 Hz and 20 kHz. A low-temperature red light LED (working voltage ~ 2 V) was used in the light-controlled capacitance measurements. The energy of the light emitted by the LED is about 1.8 eV (wave length 680 nm), which is much larger than the band gap of BST ($E_g \approx 200$ meV). [2]

## 3. Results

### 3.1. Transport measurements

Figure 1(a) shows the resistivity-temperature relation of device BST1, which indicates metallic conduction down to low temperatures, suggesting bulk-dominated conduction. By varying $V_{tg}$ at the base temperature of cryostat 1.4 K (as shown in Figure 1(b)) and sweeping a perpendicular magnetic field $B$, the Hall slope ($R_{xy}/B$), and sheet resistance $R_\square$ at 0 T, were extracted. At $V_{tg} = 0$ the total electron density is estimated to be ~ $1.6 \times 10^{19}$ cm$^{-3}$, by treating all the conduction to be bulk conduction. Given this value we conclude that $E_F$ is located in the conduction band (inset in Figure 1 (a)), as also suggested by angle resolved photoemission spectroscopy measurements. [5] The maximum applied negative voltage of $V_{tg}$ = -40 V, is close to breakdown electric field. [11] At this voltage the polarity of the Hall voltage has not yet changed, meaning that the whole material cannot be tuned from *n*-type to *p*-type. To achieve this, larger negative voltages < -60 V would be needed.

### 3.2. Capacitance measurements

Capacitance measurements were carried out on the same device. Similar results were obtained from 2 devices fabricated from the same wafer, and from another 2 devices from a different wafer but with similar Sb concentration. The inset in Figure 2(a) shows a schematic cross section view of the device. The area of the capacitor is $S = 50 \times 380$ µm² for all devices. For the measurements, coaxial wires were used to minimize the effect of parasitic capacitances. Figure 2(a) shows *C-V* traces taken at a temperature $T = 15$ K for three different frequencies ($f = 50$

Hz, 240 Hz and 20 kHz) by sweeping $V_{tg}$. At about $V_{tg} = -5$ V the capacitance signal drops precipitously and remains at the lower level for increasing negative voltage. Tuning $V_{tg}$ towards negative values causes electron depletion and the capacitance decreases as the capacitance of the depletion region gets added in series. In case an inversion layer of holes forms at the insulator/BST interface, one expects that the capacitance will stay at the low value (high frequency case) or recover its original value $C_0$ (low frequency case). Low frequency case means that at a given temperature the generation rate of holes is fast enough to follow the AC voltage modulation. In the high frequency case, in contrast, the minority carriers in the inversion layer cannot follow the external modulation $V_{AC}$. The small hysteresis observed in Figure 2(a) is characteristic for oxide charges, which are shifted back and forth by the applied gate voltage. [7] The density of this mobile charges ($n_m$) is on the order of $10^{11}$ cm$^{-2}$, which can be estimated from $n_m = C_0 \Delta V_m / e$, where $\Delta V_m$ is the voltage shift displayed in Figure 2(a). The downshift of $C_0$ with increasing frequency is the result of resistive effects. [8] For higher temperatures, here at 19.9 K (Figure 2(b)), however, the capacitance measured at $f = 50$ Hz rises again towards $C_0$, consistent with an increased generation rate (which increases with temperature) and sufficiently low frequency. At higher $T \sim 29.8$ K (Figure 2(c)) also the $C$-$V$ traces measured at higher frequencies show the characteristic low-frequency behavior, as expected for a sufficiently large generation rate. For those low frequency traces, however, they look strikingly different and a large hysteresis occurs on the back sweep. In this situation, if we pause the sweep at certain $V_{tg}$ for some time, there is no shift or drift of the curve (data not shown), suggesting the effect to be stationary. Figure 2(d) provides an overview of the temperature dependence at 50 Hz up to 118 K. High frequency $C$-$V$ traces are observed below 19.9 K, while low frequency traces are obtained when temperatures are higher. We expect that the low frequency trace survives at room temperature, since $k_\text{B}T$ ($T = 300$ K) $\approx 26$ meV is still much smaller than the band gap ($E_\text{g} \approx 200$ meV).

The evolution of the hysteresis is shown in more detail in Figure 3(a) and 3(b) in device BST2. The inset of Figure 3(a) shows the schematic "spoon" like hysteresis trace with characteristic points marked by A, B, C and D. At point A the system is in accumulation. On the way from point A to C the *C-V* trace is fully reversible, i.e. no hysteresis occurs (apart from the small hysteresis ascribed to moving charges in the oxide layer). Once point C is reached, however, the pronounced hysteresis starts to develop. The evolution of the hysteresis from the inversion side is shown in Figure 3(b). Here, hysteresis starts to occurs once $V_\text{tg}$ is swept beyond point C towards larger gate voltages. Similar pronounced hystereses had been found in X-ray irradiated (applied to increase the conventional surface state density) silicon based MOS capacitors, as well as in silicon-carbide gate-controlled diodes. [12, 13] There, the origin of the hysteresis was ascribed to the existence of deep-lying (trapped) surface states. These states, sufficiently far away from the valence and conduction band edge can be charged/discharged by capture of holes/electrons when $E_\text{F}$ is close to the valence band ($E_\text{v}$)/the conduction band ($E_\text{c}$).

### 3.3. Discussion on the "spoon" like hysteresis

Given the non-perfect BST/insulator interface, it is obvious to assume that besides delocalized topological surface states (TSS), localized surface states (LSS) also exist. These trap states are energetically distributed across the bandgap (see Figure 4) and are likely the origin of the observed hysteresis. In the framework of carrier generation and recombination theory, the capture rate of carriers into surface states is proportional to carrier concentration while the emission rate from localized states with energy $E_\text{T}$ decreases with increasing $E_\text{c} - E_\text{T}$ ($E_\text{T} - E_\text{v}$). [12, 14] As we lack detailed information about the nature of the defect states we follow the reference, [12] to explain the *C-V* traces: At accumulation (point A, Figure 4(b)), all the surface states are charged with electrons. Between points A and B (Figure 4(c)), traps are still able to emit electrons and to follow the changing bias voltage $V_{tg}$. From point B to C (Figure 4(d)), most surface states stay charged negatively and can no longer follow the applied bias. Sweeping further towards point D, the surface states get gradually compensated, due to the appearance of holes under inversion. This is associated with a change of slope at point C. In case $V_{tg}$ is swept to even more negative values, the capacitance approaches $C_0$ (see inset Figure 4(a)). At this point all the negative charge has been annihilated and the trap states above $E_F$ are filled with holes (Figure 4(g)). But most of the time we have avoided sweeping to these large negative voltages, as they caused enhanced hysteresis. Usually we swept back at a point D before reaching full inversion. By moving back from point D to E (Figure 4(e, f)), due to the low concentration of electrons, the empty trapped states stay discharged serving as an extra space charged region, inducing lower total capacitance. Upon reaching point B all the trapped states will be charged instantly with electrons up to $E_F$, leading to a sharp increase of capacitance; the system recovers to equilibrium and the hysteresis disappears.

From the data shown in the inset of Figure 4(a), the trapped density of surface states ($\Delta n$) between points C and D' can be estimated. For the voltage interval $\Delta V \sim 11.8\ V$, $\Delta n = C_0 \Delta V/(Se) \approx 6 \times 10^{12}$ cm$^{-2}$. We note that the density of TSS within the band gap is smaller but of the same order of magnitude compared with $\Delta n$, thus we cannot exclude that TSS also contribute to the hysteretic behavior. In this case, however, it would be unclear how charge can be trapped in these metallic like delocalized TSS.

### 3.4. Effects of light on the capacitance

It is expected that by introducing external electron and hole pairs, for example, by illumination, the C-V traces will be strongly influenced.[15] This also applies in our devices. The experiments of illumination on the C-V traces were done on device BST3 with carrier density ($V_{tg} = 0$) of $6.7 \times 10^{18}$ cm$^{-3}$ at 1.4 K, as estimated from the Hall slope. In contrast to the devices discussed above, it was protected by a few nm Al$_2$O$_3$ obtained by oxidation of a 0.7 nm (nominal) in-situ capped Al layer after BST growth, since on one hand, we want to see if the variation of insulator plays a role in the hysteresis, and on the other hand it may improve the surface quality. A similar approach of protecting the TI surfaces has been used by other groups before. [16] Corresponding data are shown in Figure 5 for two temperatures, where we observed transition between high and low C-V traces. When no light is applied, compared with devices BST1 and BST2, the change of insulator shifts the flat band voltage (for example, ~-1 V for BST2, and ~-7.5 V for BST3), due to the different dielectric constants of the insulators, as well as the different doping levels of the surface states of the device during growth and device fabrication. However, similar hysteresis also presents, suggesting that the trapped states are already formed during the BST growth. In addition, the C-V traces change drastically upon illumination. While at 10 K, high frequency C-V curves prevail, a low-frequency C-V trace with large hysteresis dominates at 20 K in the dark. In the latter case the C-V traces become essentially "flat" at the highest LED intensities. This is expected for metal like systems with high carrier concentration which mimics a capacitor with two metal plates and capacitance $C_0$.

4. Conclusions

We observed typical field effect C-V traces in compensated n-type (Bi$_{1-x}$Sb$_x$)$_2$Te$_3$ topological insulator films, showing accumulation, depletion and inversion. As in MOS devices the inversion capacitance strongly depends on temperature, AC frequency and illumination. A strong hysteresis observed under inversion points to the presence of a high density of trapped surface states ($6\times10^{12}$ cm$^{-2}$) coexisting with topological surface states. The possibility to tune not only the density of topological surface states by the top gate voltage, but also the bulk surface charge adds a new option to manipulate topological surface states.

## Acknowledgements

The authors greatly appreciate the financial support from Deutsche Forschungsgemeinschaft via SFB1277 (A08) and the Alexander von Humboldt Foundation. This project has received funding from the European Research Council (ERC) under the European Union's Horizon 2020 research and innovation programme (grant agreement No 787515).

Figure Captions

**Figure. 1**    Transport properties of BST1. (a) Resistivity-temperature dependence. The upper inset sketches the band structure with Fermi level position, the lower one shows an optical image of the device. (b) $V_{tg}$ dependence of $R_{xy}/B$ and zero magnetic field $R_\square$ at 1.4 K. The horizontal blue line marks $R_{xy}/B = 0$, at which the total carrier changes type. The dashed blue line extrapolates to the gate voltage at which this would happen.

**Figure. 2**    C-V traces at different frequencies and temperatures. (a-c) C-V traces taken at 3 different frequencies (50 Hz, 240 Hz, 20 kHz) at (a) 15 K, (b) 19.9 K, and (c) 29.8 K, respectively. Inset in (a) shows a schematic side view of the measurement setup. Dotted arrows in (a-c) indicate the sweep directions. Dashed lines and arrows in (b) approximately mark the regions of accumulation, depletion and inversion. (d) temperature dependent C-V traces taken at 50 Hz. The increased saturation capacitance value at positive $V_{tg}$ may be attributed to an increased parasitic capacitance when the temperature is higher. In addition, the $V_{tg}$ needed for the capacitance to drop (depletion) moves to the negative side, suggesting an increased surface carrier density which needs to be depleted, when the temperature increases. Since depletion and inversion are observed all the time, we conclude that the top surface area can be tuned from *n* to *p* type at all temperatures, the same as that at 1.4 K. All the C-V traces were taken with the same $V_{tg}$ sweeping sequence: 0 to +5 V to -15 V to 0 in 0.2 V steps and 0.1 V/s sweeping rate. The time intervals between adjacent points are about 45 s, 20 s, and 12 s for 50 Hz, 240 Hz, and 20 kHz, respectively.

**Figure. 3** Minor hysteresis loops of device BST2 starting from (a) accumulation and (b) inversion. Each line or loop is shifted by -0.1 pF along the y axis for clarity. The dotted arrows specify the sweep direction of the loops, and blue arrows mark the gate voltage at which the loops start to open. The inset in (a) shows a schematic hysteresis loop with characteristic points A to D discussed in the text.

**Figure 4.** Possible explanation on how the hystereses form. (a) $C$-$V$ hysteresis loop of device BST1 at $f = 50$ Hz and $T = 29.8$ K between $V_{tg} = +5$ V and -15 V. The inset shows data from the same device but swept to -25 V. At this large negative voltage strong hysteresis occurs and the capacitance increases towards $C_0$. (b-g) Schematic band bending upon sweeping from accumulation towards inversion and back corresponding to points A, B, C, D, and E, respectively in (a), as well as point D' in the inset of (a). The filling of the LSS is sketched on the left hand side of each figure (blue area); topological surface states are not shown. In case (b) and (c) the Fermi level in the surface states is sufficiently close to the conduction band edge, so that the filling of the surface states follows the gate voltage. In (d) a significant amount of negative charge (blue area above $E_F$) is already trapped as the emission rate is small. At about this point annihilation of electrons by holes starts to occur. In (e) the trapped states become partly discharged (blue area above $E_F$). Upon sweeping back to depletion, the partly empty trapped states stay (in (f)) due to the low capture rate of electrons, until (c) is reached, where they will be filled up with electrons instantly. (g) is the transition point where $E_F$ is so close the valence band edge, that all trapped states are filled with holes.

**Figure. 5**   *C-V* traces for different LED intensities (given by the bias current, the working voltage is almost constantly 2 V) at (a) 10 K and (b) 20 K in device BST3.

Figures

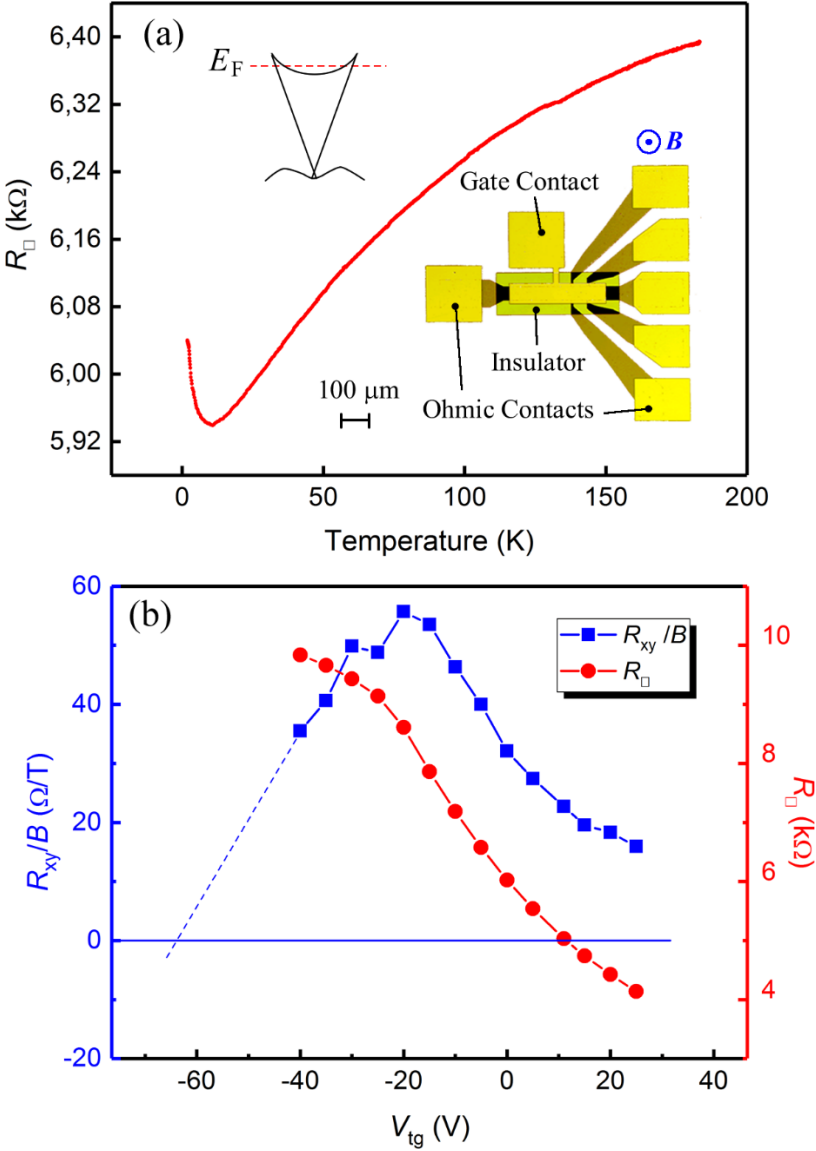

Figure 1

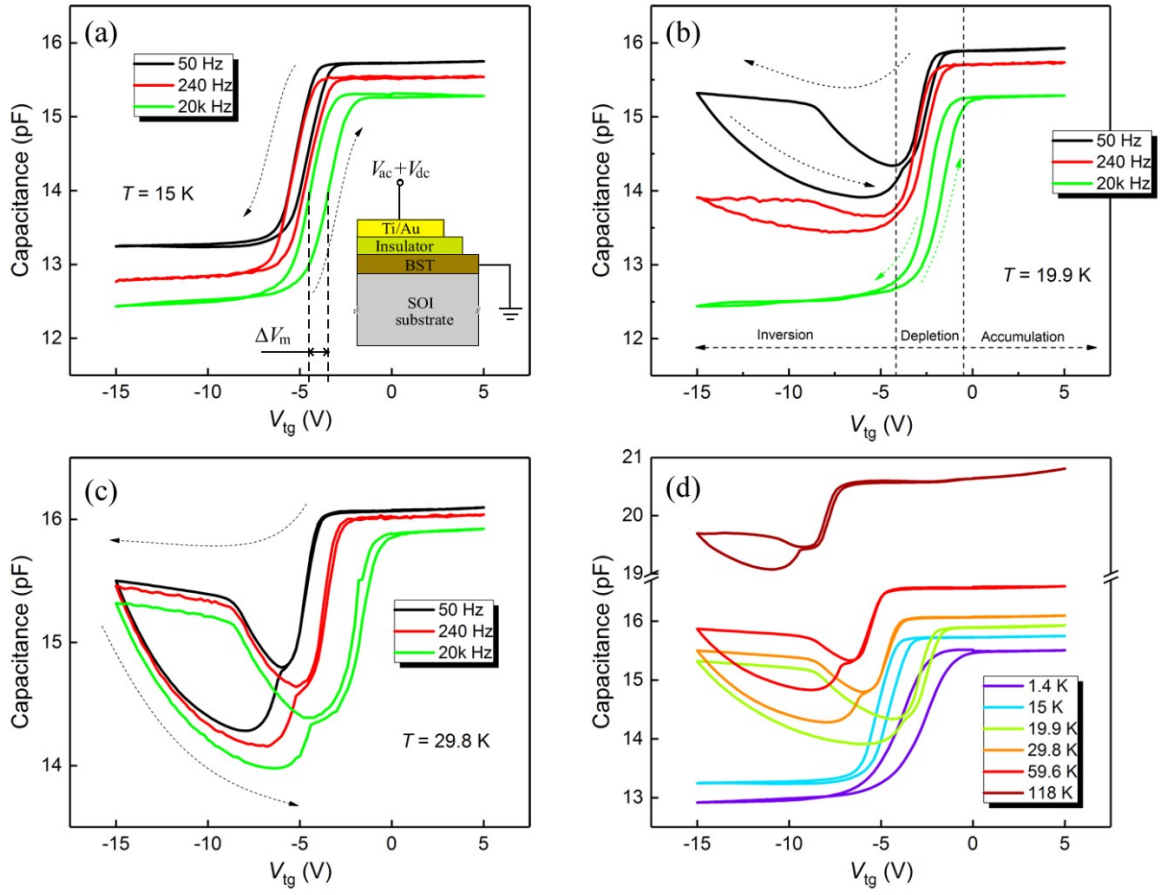

Figure 2

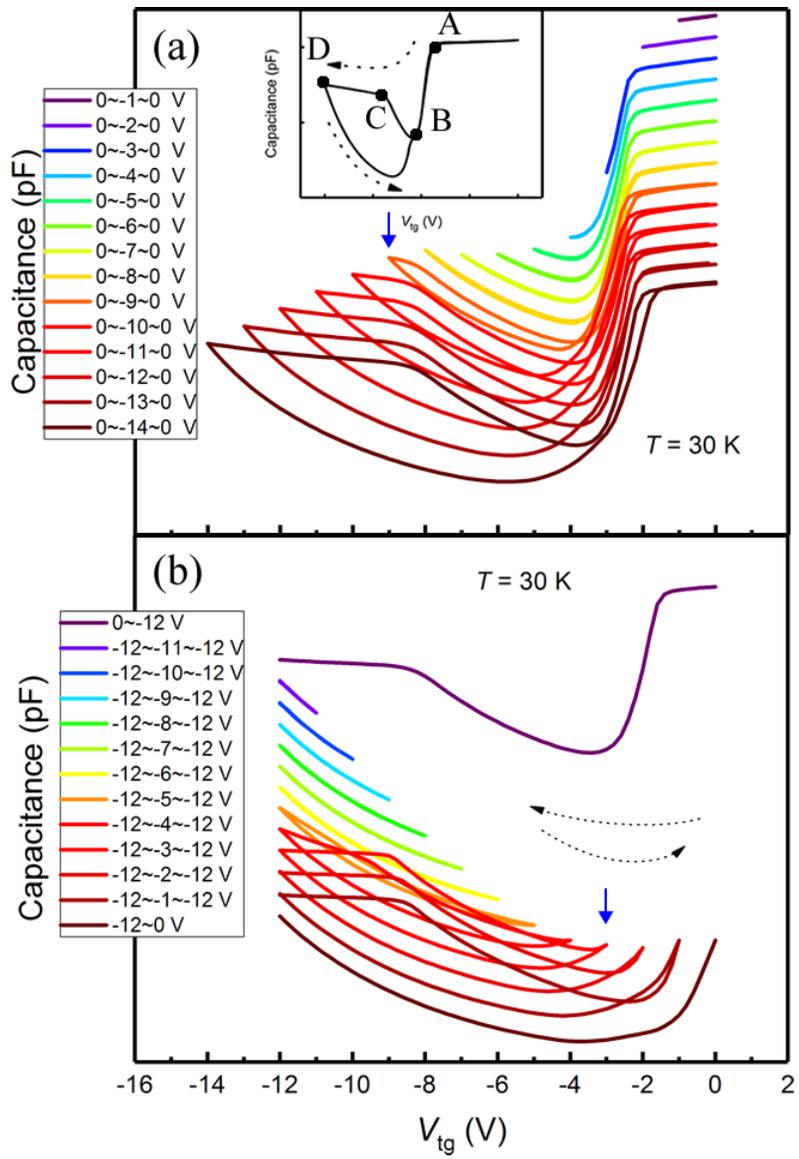

Figure 3

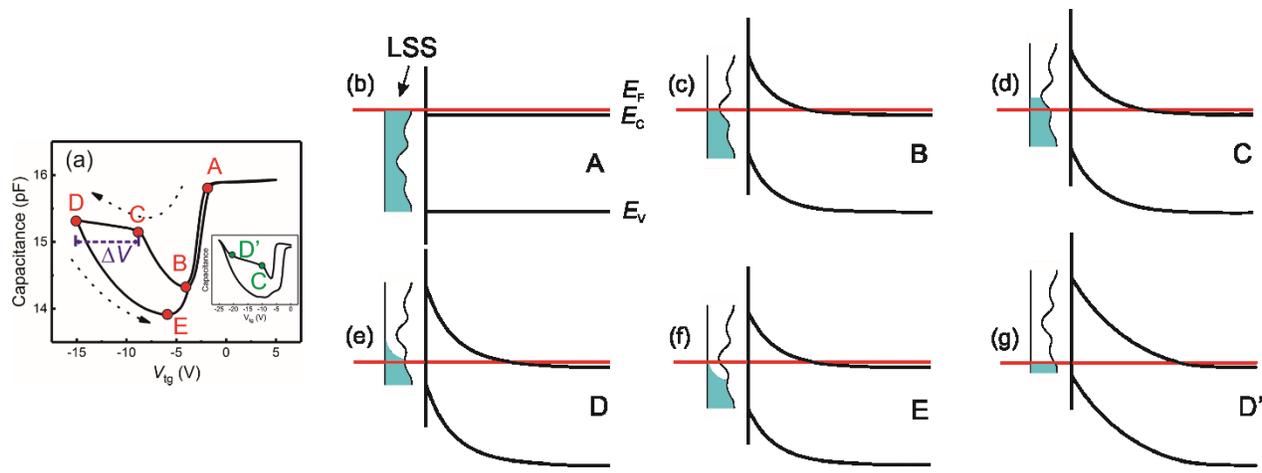

Figure 4

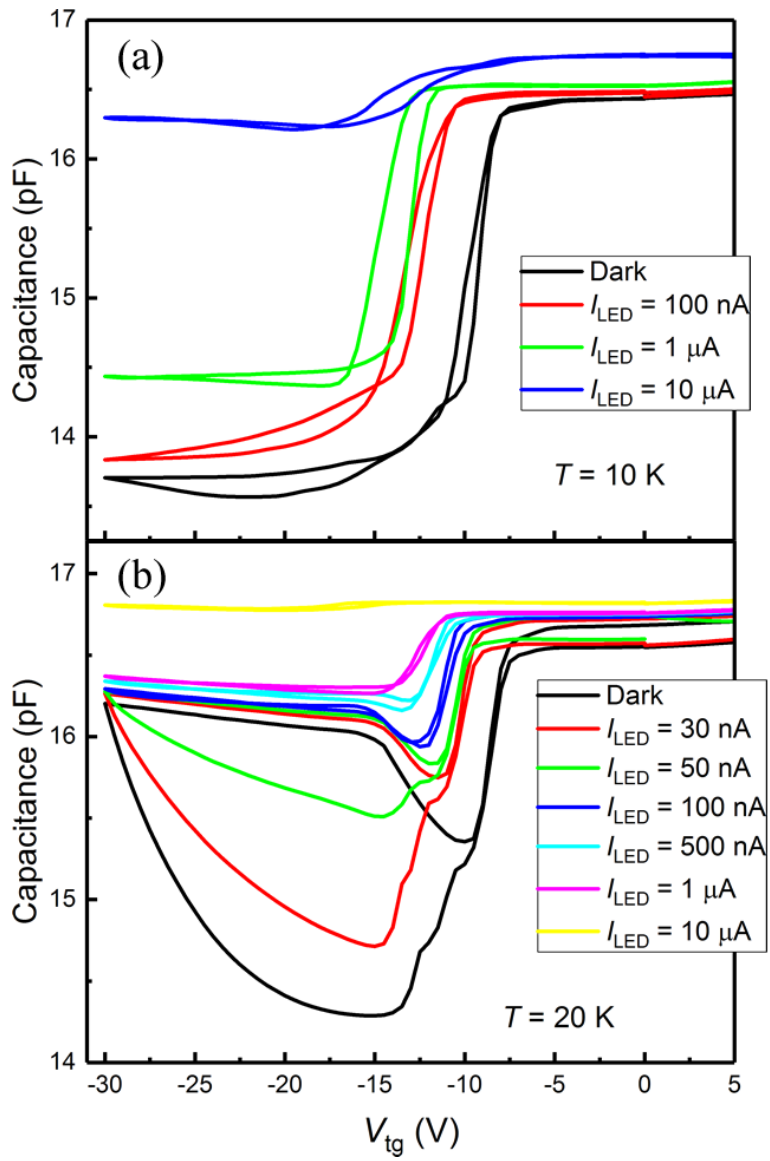

Figure 5